\documentstyle{mn}

\def\etal{{et al.\ }}

\def\xstar{{\sc xstar~}}

\begin{document}

\title{\bf The broad-band X-ray spectrum of Mrk 3}

\author[R. G. Griffiths et al.]{R. G. Griffiths$^{1}$,  R. S. Warwick$^{1}$, 
I. Georgantopoulos$^{1,2}$, C. Done$^{3}$ and
\newauthor
D. A. Smith$^{1}$  \\ 
$^1$Department of Physics and Astronomy, University of Leicester,  
Leicester, LE1 7RH \\
$^2$Astronomical Institute, National Observatory of Athens, Lofos Koufou,
Palaia Penteli, Athens, GR - 15236, Greece \\
$^3$Department of Physics, University of Durham, South Road, Durham DH1 3LE} 

\maketitle

\begin{abstract}
We have used {\it non-simultaneous} {\sl Ginga}, {\sl ASCA} and 
{\sl ROSAT} observations to investigate the complex X-ray spectrum of 
the Seyfert 2 galaxy Mrk 3.  We find that the composite spectrum
can be well described in terms of a heavily cut-off hard X-ray 
continuum, iron $K_{\alpha}$ emission and a soft X-ray excess, with
spectral variability confined to changes in the continuum 
normalisation and the flux in the iron line.
Previous studies have suggested that the power-law continuum in 
Mrk 3 is unusually hard. We obtain a canonical value for the
energy index of the continuum (i.e. $\alpha \approx 0.7$) when a
warm absorber (responsible for an absorption edge observed near 8 keV) 
is included in the spectral model. Alternatively, the inclusion of a 
reflection component yields a comparable power-law index.   
 The soft-excess flux cannot be modelled solely in terms 
of pure electron scattering of the underlying power-law continuum.
However, a better fit to the spectral data is obtained if we include
the effects of both emission and absorption in a partially photoionized
scattering medium.  In particular the spectral feature prominent at
$\sim 0.9$ keV could represent O VIII recombination radiation 
produced in a hot photoionized medium. We discuss our results in
the context of other recent studies of the soft X-ray spectra of Seyfert 2
galaxies.
 
\end{abstract}


\newpage

\section{INTRODUCTION}

Seyfert galaxies have been observed extensively in the high energy band to 
reveal a host of interesting properties. Seyfert 1 galaxies are typically 
much brighter X-ray sources than Seyfert 2's, a result which is explained by 
current unification models as primarily due to the effects of obscuration and 
orientation (e.g. Antonucci 1993). Both types of object are presumed to contain
an active galactic nucleus (AGN) comprising a supermassive blackhole, an 
accretion disk, a broad-line region and (in many cases) a thick molecular 
torus. According to the unified scheme, we have a relatively clear view of 
the AGN in a Seyfert 1 galaxy, whereas in Seyfert 2 galaxies the torus is 
observed at sufficiently high inclination so as to introduce significant
obscuration into the direct line-of-sight to the central source. Recent X-ray 
spectral observations provide clear evidence for highly ionized absorption 
systems (i.e. warm absorbers) in many Seyfert 1 galaxies  
(Reynolds 1997; George et al. 1998). It appears that this ionized material 
extends sufficiently far from the nucleus along the axial direction, so as to 
be visible above the torus even when the system is viewed close to edge-on. 
In Seyfert 2's this medium can scatter nuclear flux into our line-of-sight, 
even when our direct view of the nucleus is entirely blocked by the torus 
(e.g. NGC 1068, Antonucci \& Miller 1985).  

As a result of missions such as {\sl Ginga}, {\sl ROSAT} and {\sl ASCA} and,
more recently, {\sl RXTE} and {\sl SAX}, Seyfert 1's and Seyfert 2's have been
shown to exhibit a rich phenomenology in terms of their X-ray spectra.  
Recent interest has centred on (i) the temporal and spectral properties of 
the hard X-ray continuum produced close to the supermassive blackhole; (ii) 
the presence of spectral components resulting from the Compton reflection of 
continuum flux by optically thick matter in the proximity of the central 
source; (iii) the detailed study of the iron K$_{\alpha}$ line (in particular 
the presence of broad-line profiles commensurate with predicted gravitational 
redshifts and Doppler broadening effects) ; (iv) the presence of highly 
ionized material whether viewed directly against the continuum source as a 
`warm-absorber' or, indirectly, via scattered nuclear emission in those 
sources where the direct soft X-ray continuum is strongly cut-off due to 
line-of-sight absorption.

In the present paper we focus on the X-ray properties of the Seyfert 2
galaxy, Mrk 3. Our approach is to use {\it non-simultaneous}  {\sl Ginga}, 
{\sl ASCA} and {\sl ROSAT} observations to secure broad (0.1--30 keV) 
spectral coverage. Although Mrk 3 has previously been studied in some detail 
on the basis of the individual datasets (Awaki et al. 1990;
Awaki et al. 1991; Iwasawa et al. 1994; Turner, Urry \& Mushotzky 1993;
Smith \& Done 1996; Turner et al. 1997a,b),  the combination of the data 
from the three satellites provides clear benefits in terms of spectral 
modelling. For example, the {\sl ASCA} instruments, despite their superior 
spectral resolution, do not have the spectral bandwidth to constrain the 
slope of the underlying hard X-ray continuum in a source, such as Mrk 3,
where this continuum is strongly cut-off in the soft X-ray regime. However, 
once the continuum form is 
reasonably specified by the inclusion of {\sl Ginga} data, then the {\sl ASCA} 
observations provide tight constraints on sharp emission features. Similarly 
the inclusion of {\sl ROSAT} data extends the spectral coverage below 
$\sim 0.6$ keV to give additional leverage when modelling the soft X-ray 
spectral form. The consideration of the composite spectrum also allows the 
possibility of spectral variability to be investigated in a consistent manner. 

One particular point of interest in the case of Mrk 3 is that
it appears to have a rather hard continuum form. The {\sl  Ginga} 
observations  (Awaki \etal   1990; Awaki \etal 1991) reveal a power-law 
index $\alpha \approx 0.5$, which is somewhat harder than is typical
of Seyfert 1 galaxies (e.g.  Nandra \& Pounds 1994) and fairly extreme 
amongst the sample of $\sim 16$ Seyfert 2's studied by {\sl Ginga} 
(Smith \& Done 1996). In principle the observation of anomalously flat 
power-law slopes, even in a subset of Seyfert 2 galaxies (NGC5252 provides a 
further example; Cappi et al. 1996) could be interpreted as contradicting 
unification models. Flat spectrum populations are also of general interest 
in terms of their potential contribution to the cosmic X-ray background (XRB). 

This paper is organised as follows. Brief details of the observations and 
data reduction methods are presented in section 2.  In the next section we 
describe the approach we have taken in modelling the various spectral 
datasets, including the use of the photoionization code \xstar (Kallman \& 
Krolik 1997) to calculate the emergent spectrum from a partially photoionized 
scattering medium.  In section 4 we discuss the implications of our results
in the context of other, recently published studies of the X-ray spectral
properties of Seyfert 2 galaxies. Finally we summarise our conclusions
in section 5.

\section{OBSERVATIONS AND DATA REDUCTION}

\subsection{The {\sl Ginga} observations}

Mrk 3 was observed with the Large Area Counter (LAC,
Turner \etal 1989) on board the {\sl Ginga} satellite (Makino \etal
1987) during the period September 27-28, 1989. The methods employed in the 
data selection, in the background subtraction and in correcting for pointing 
offsets were the same as those described in Smith \& Done (1996).
X-ray pulse-height spectra were extracted from {\it both} the ``top-'' 
and ``mid-layers'' of the LAC; these counters provide spectral coverage 
across nominal 2--37 keV and 6--37 keV energy bands respectively. 
The effective integration time was 24 ks and the average count
rates were $4.6 \rm~count~s^{-1}$ and $2.4 \rm~count~s^{-1}$ in the top
and mid-layer counters respectively. In the spectral analysis,  
data below 4.5 keV were ignored because of the presence of anomalous line 
features at $\sim$ 3--4 keV (possibly identified with argon K and xenon L 
lines in the LAC background; see Smith $\&$ Done 1996) and high energy 
channels above 27 keV were excluded due to their susceptibility to 
background subtraction errors. A one percent systematic error was added to 
each pulse-height channel to account for uncertainties in the detector 
response.

\subsection{The {\sl ASCA} observations}

{\sl ASCA} observed Mrk 3 during the period April 21-22, 1993. There are 
four instruments on-board {\sl ASCA}, two Solid-State Imaging Spectrometers 
(SIS-0 and SIS-1) 
and two Gas Imaging Spectrometers (GIS-2 and GIS-3) (Tanaka, Inoue \& Holt
1994). The SIS data were taken in 4-CCD FAINT and BRIGHT modes.  
We combined FAINT mode data (after conversion to BRIGHT 
mode), and true BRIGHT mode data.  This procedure does not exploit the 
maximum energy resolution of the SIS, but it does increase the signal-to-noise 
ratio. (We obtained $\sim$ 40$\%$ more counts by using the combined true and 
converted BRIGHT mode data rather than the FAINT mode data alone). 
The events lists have been screened 
using the standard  {\sl ASCA} analysis procedure. Only events of grade 
0, 2, 3 and 4 have been used and data taken during intervals when the Earth 
elevation angle was less than $5^{\circ}$ and when the magnetic cut-off 
rigidity was less than $6\rm~GeV/c $ were rejected. For the SIS data, ``hot''
and ``flickering'' pixels and telemetry saturated frames were also
removed. In addition, 
counts were rejected when the bright Earth angle was less than 
$20^{\circ}$.  This procedure resulted in 30 (42) ks of good SIS (GIS) 
data for Mrk 3. 

Spectral data sets were obtained from each telescope by integrating events 
within a circular region (of radius 3--4 arcmin) centred on Mrk 3 and 
the background was estimated from source-free regions in the same
observation. The average background-subtracted count rates for Mrk 3
were $0.05 \rm~count~s^{-1}$ and $0.04 \rm~count~s^{-1}$ in the SIS
and GIS systems respectively. 
In the subsequent spectral analysis we have ignored all 
SIS (GIS) data below 0.6 (0.8) keV (since the uncertainty in the calibration
is highest in these low energy channnels). The pulse-height spectra were
binned so as to give at least 20 counts per spectral channel and, in the 
spectral analysis, the data from all four detectors were fitted simultaneously.

\subsection{The {\sl ROSAT}/PSPC observations}

Mrk 3 was observed with the X-ray Telescope (XRT) and 
Position Sensitive Proportional Counter (PSPC) combination on board 
{\sl ROSAT} during the period March 8-10, 1993. 
Data were rejected when the Master Veto rate exceeded a 
threshold of $170 \rm~ count~s^{-1}$, which resulted in a total exposure 
time of $\sim$ 15 ks. The
background region was taken from an annulus with an inner radius
of 9$^{\prime}$ and outer radius of 15$^{\prime}$, centred on Mrk 3, with 
all the confusing sources masked out. The observed 0.1--2.4 keV PSPC count 
rate was  $0.06 \rm~count~s^{-1}$.

\section{Spectral Analysis}

In this section we first consider the {\sl Ginga} spectra and 
{\sl ASCA} spectra separately so as to allow direct comparison with 
earlier studies. We then go on to apply more detailed spectral models 
to the composite  {\sl Ginga}, {\sl ASCA} and {\sl ROSAT} spectra. 
The spectral analysis was conducted using XSPEC version 10.0. 

We note that there is a faint 
X-ray source detected $\sim 1.7^{\prime}$ West of Mrk 3 in
both the ROSAT PSPC (Turner et al. 1993) and  HRI (Morse et al. 1995)
images.  The source flux is only $\sim$ 4.4 $\%$ of the flux from 
Mrk 3 in the 0.1--2.4 keV PSPC band.  The presence of this faint source 
within the {\it Ginga} LAC field of view and in the {\it ASCA}
``on-source'' region  is very unlikely to produce a significant
effect in terms of our analysis of the Mrk 3 spectrum. 
(Furthermore the centroid of the source lies in the interchip 
gap for both the SIS0 and SIS1 detectors). Iwasawa et al. (1994)
and Turner et al. (1997a) also report the detection of a BL Lac object,
$10.6'$ NW of Mrk 3 in the {\sl ASCA} data; this source also lies within 
the {\sl Ginga} field of view. Based on the {\sl ASCA}
data we estimated that the count rate from the BL Lac object is roughly 2\% 
of that of Mrk 3 in the 4--10 keV band and hence the impact of this confusing 
source on the {\sl Ginga} measurements will be small.

\subsection{The {\sl Ginga} spectra}

The first model used to fit the {\sl Ginga} LAC spectra comprised a single 
hard power-law continuum, with energy index $\alpha$ and normalisation 
$A_{Ginga}$, which is heavily cut-off at low energy due to
absorption by cold solar abundance gas of column density, $N_{H}$.
(Throughout this paper we have used the element abundances and the 
absorption cross-sections adopted by Morrison \& McCammon 1983).
An iron K$_{\alpha}$ emission line of intensity 
$A_{K\alpha}$ at an energy $E_{K\alpha}$ was also included with a
width fixed at $\sigma_{K\alpha}$ = 0.1 keV (we use this value 
throughout, see section 4).
This model (Model 1) provides a good match to the {\sl Ginga} data for 
Mrk 3;  Table 1 lists the best-fit parameters and 
associated errors (the errors quoted throughout this paper are 90 per cent 
confidence limits for one interesting parameter following the prescription 
of Lampton, Margon \& Bowyer 1976).  

The parameter values derived on the basis of Model 1 agree reasonably well 
with earlier analyses of the {\sl Ginga} (top-layer only) spectra 
(Awaki et al. 1990;  Awaki \etal 1991). Specifically we confirm, 
on the basis of the {\sl Ginga} data alone, the rather hard continuum 
spectrum (i.e. $\alpha = 0.37^{+0.15}_{-0.14}$). However, 
one apparent discrepancy is that Smith \& Done (1996) obtained 
$\alpha = -0.15^{+0.13}_{-0.19}$ for Mrk 3, which is a very much flatter 
continuum slope than quoted here. The origin of this problem can be traced
to the fact that Smith \& Done restricted their analysis to the 2--18 keV 
energy range whereas we use the 4.5--27 keV band.  If we employ Model 1 
to fit the Ginga 2--18 keV top-layer data for Mrk 3, then we also
obtain a peculiar spectral index. This emphasises the value 
of the high energy data, particularly that from the LAC mid-layer.
(The mid-layer has a larger effective area than the top-layer above 
$\sim 10$~keV and therefore provides useful additional 
constraints on the slope of the hard X-ray continuum).

The measured {\sl Ginga} count rate spectra, the best-fitting spectral model 
and the spectral fitting residuals are shown in 
Fig. 1. The residuals suggest the possible presence of an additional 
absorption edge at $\sim 8$ keV. When such a  feature is 
included  a reduction in $\chi^{2}$ of $\sim 14$ is 
obtained with two additional parameters, the edge energy $E_{edge}$ and 
optical depth $\tau_{edge}$ (Model 2, Table 1), which in terms of the F-test 
is significant at the $>$ 99 $\%$ level. One effect of this additional 
iron absorption is to steepen the
continuum by $\Delta\alpha \approx 0.3$.  The measured edge energy, 
$E_{edge} = 8.1^{+0.4}_{-0.5}$ keV, is consistent with the K-edge of iron
in an ionization state in the range Fe XVIII-XXIII, implying the presence
of highly ionized gas in the line-of-sight to the hard X-ray source
in addition to the large column density of relatively cold gas.  
The possibility of additional line-of-sight absorption by photoionized gas is 
discussed further in section 3.3.  

\subsection{The {\sl ASCA} spectra}

Mrk 3 exhibits excess soft X-ray emission over and
above that expected given the heavy absorption evident in the 
{\sl Ginga} passband. To account for this soft X-ray emission the spectral 
form used to fit the {\sl ASCA} SIS and GIS data was basically  
that of Model 1 (with free parameters $\alpha$, $A_{ASCA}$, $N_{H}$, 
$A_{K\alpha}$, $E_{K\alpha}$) plus a second power-law with the same 
spectral slope as the intrinsic hard continuum but with a different 
normalisation ($A_{soft}$). Hereafter we will refer to this
as the soft power-law component. In physical terms this soft power-law
could be flux leakage from a partially covered source
with an {\it uncovered} fraction $f = A_{soft}$/$A_{ASCA}$. Alternatively
$f$ could represent the fraction of the continuum flux which is directed into 
our line of sight by an extended electron scattering region 
(i.e. $f \equiv f_{scat}$). We will, in fact, pursue this latter 
interpretation in the current paper. We assumed that the soft power-law 
component is absorbed only by the Galactic foreground HI column density 
of $N_{gal} = 8.7 \times 10^{20} \rm~cm^{-2}$ (Elvis, Lockman \& Wilkes
1989).

The above spectral prescription provides only a modestly successful fit to 
the {\sl ASCA} 
data (Model 3, Table 2).  Significant residuals are evident at a variety 
of energies in the {\sl ASCA} detectors. Previously Iwasawa et al. (1994) 
have reported the presence of line-like features at $\sim$ 0.75 keV, 
0.9 keV and possibly $\sim 7.0$ keV in the SIS spectra of Mrk 3 (in addition
to the 6.4 keV line included in Model 3). Similarly  
Turner et al. (1997a) found that the addition of narrow lines at 1.85, 2.45, 
6.4 and 6.96 keV (corresponding to SiXIII, SXV, FeI-XVI and FeXXVI  
respectively) to their  best-fitting double power-law continuum model, 
significantly improved the fit to the ASCA data. 
The significance of any residual features is, of course, a strong function 
of the continuum model fitted (see Turner et al. 1997a).

We obtain an improved fit if we include a solar-abundance Raymond-Smith 
component (temperature $kT_{Ray}$ and 
normalisation $A_{Ray}$), {\it in addition to} the soft power-law
(Model 4, Table 2). One aspect to note is 
that Model 4 gives a much flatter power-law slope than Model 3. The 
difference appears to arise from the fact that in Model 3  the intrinsic and 
soft power-laws are constrained to have the same spectral index and the 
steepness of the 
soft excess tends to pull the overall continuum form to a steeper slope,
whereas when the Raymond-Smith component is included this linkage
is much weaker. However, the soft X-ray power-law component is still
essential since fits of Model 4 with $A_{soft}$ set to zero
give a worse result than even Model 3 ($\chi^{2}$/d.o.f. = 443/299 
compared to $\chi^{2}$/d.o.f. = 390/300).

In the next section we focus our attention on the possibility 
of fitting some of the line-like features with a  model which includes
the effects of  photoionization on the putative scattering medium
(and hence removes the requirement for a Raymond-Smith component).
For completeness we include the results of fitting such a model to the 
{\sl ASCA}-only data as Model 5 in Table 3 (details of this model are 
explained in the next section). The fact that this photoionization
model gives a rather similar $\chi^{2}$ to Model 4 suggests that it is
difficult to distinguish between the two descriptions at the limits
of the sensitivity and spectral resolution afforded by {\sl ASCA}.

\subsection{ The composite {\sl Ginga}, {\sl ASCA} and {\sl ROSAT} spectra}

Since the {\sl Ginga}, {\sl ASCA} and {\sl ROSAT} observations were not 
recorded contemporaneously  (the observations were, in fact, made over a 
period of $\sim 4$ years) our analysis is based on the 
assumption that the spectral form of Mrk 3 is largely invariant.
Thus in the spectral analysis we allow only the normalisation of the 
intrinsic power-law continuum and the flux of the iron K$_{\alpha}$ line
to vary between the {\sl Ginga} and {\sl ASCA} datasets; the spectral 
parameters used to model the {\sl ROSAT} PSPC spectrum were identical to those 
applied to the {\sl ASCA} data. 
We will check this assumption later
when considering the quality of the model fit to the composite spectra. 

Table 3 gives the results of fitting Models 3 and 4  to the composite spectra
for Mrk 3. The results generally mirror those obtained from the ASCA spectra 
alone, although as noted earlier the Ginga data prefer a rather harder power 
law and this influence is evident from a comparison
of the results in Table 3 with those in Table 2. By way of
illustration Fig. 2 shows the unfolded spectral data based on Model
3 and the corresponding count rate residuals to the composite 
spectrum. Interestingly the residual feature near 0.9 keV identified by
Isawawa et al. (1994) appears to be present both in the {\sl ASCA} and 
{\sl ROSAT} data. This is also consistent with the report by 
Turner et al. (1993) of the presence a blend of soft X-ray lines below
$\sim 1$ keV in the {\sl ROSAT} PSPC spectrum of Mrk 3.
Other distinctive line features are less obvious. For example, both
Iwasawa et al. (1994) and Turner et al. (1997a) report the presence of a 
line near 7.0 keV which could be K$_{\alpha}$ emission from hydrogen-like 
iron. We have investigated whether a narrow line is present at 6.87 keV (i.e.
the redshifted energy of the Fe XXVI line) but find that this line is at best
$\sim$ 20 times weaker than the 6.4 keV line and that its inclusion
gives no significant improvement in the spectral fitting. Iwasawa et al. 
(1994) note that in their analysis this line is only marginally detected. 
We conclude that this is a case where the continuum model has a crucial 
bearing on the inferred significance of a line feature. 

The inclusion of the solar-abundance Raymond-Smith component in addition to
the soft power-law continuum (Model 4) provides a dramatic improvement in the
fit to the soft X-ray spectrum of Mrk 3. Figure 3(a) illustrates the form of
this spectral model in the 0.3--3 keV band. With the {\sl ASCA} spectral
resolution, the line emission from the $\sim 0.7$ keV thermal component gives
rise to a broad spectral enhancement in the range 0.7--1.0 keV in rough accord 
with that observed.  As a trial we reduced the elemental abundance in the
Raymond-Smith model to 0.2 solar and refit the spectra.  The effect was to
increase the contribution of the thermal continuum to the soft X-ray spectrum,
particularly below 0.6 keV.  As a result, the slope of the power-law continuum
flattens somewhat from $\alpha \approx 0.45$ to $\alpha \approx 0.37$ giving a
slightly improved fit ($\Delta \chi^{2} \sim 7$).

As discussed earlier we can interpret the soft power-law component in 
Models 3 and 4 as the fraction of the intrinsic hard X-ray continuum 
which is electron scattered into our line of sight. Unless the scattering
region is completely photoionized by the incident nuclear flux, we would 
expect some absorption features to be imprinted on the scattered flux and 
also some line emission to emanate from the photoionized medium.  Thus we 
now focus our attention on the possibility of extending Model 3
to include both absorption features in the scattered continuum
and intrinsic emission from the scattering region.

We have used the photoionization code {\sc xstar\/} (Kallman \& Krolik 1997)
which models the emission and absorption of a cloud of photoionized gas.
This photoionization code calculates the emission and
absorption spectra produced by a gas cloud in the vicinity of a point
source of continuum radiation.  The assumed spectrum of the central 
ionizing source
is the same one used by Krolik \& Kriss (1995) and is typical of Seyfert 1
galaxies and low-luminosity quasars.  The state of the gas cloud is described
by the ionization parameter $\xi$ = L/{\it n}r$^{2}$ where L is the
source luminosity in the 0.0136 -- 13.6 keV bandpass 
in erg s$^{-1}$, {\it n} is number of hydrogen atoms/ions
in the gas per cm$^{3}$ and r is the distance of the inner edge of the
cloud to the central source in cm. 
In our {\sc xstar} models we have used a density of {\it n} =
10$^{6}$ cm$^{-3}$ and a gas temperature of 10$^{6}$ K (but see section 4.2).
The element abundances used are the default values in {\sc xstar\/}. 
A range of ionization parameters (log $\xi$) and column densities 
($N_{H_{scat}}$) were used, from which a two-dimensional grid of spectral 
models was produced in the form of an ``atable'' and ``mtable''.  The atable
represents the intrinsic emission of the photoionized plasma, and the mtable
specifies the fraction of the scattered power-law which emerges, after
allowing for absorption, from the cloud. In both cases the tabulated spectral 
information encompassed an observational bandwidth from 0.1 keV up to 40 keV.
In the spectral fitting XSPEC interpolates between the grid points to 
obtain the best-fitting spectral parameters.

We assume that the absorption suffered by the scattered continuum is that
arising in a slab of gas directly illuminated by the nuclear source in 
Mrk 3.  The intrinsic emission from the scattering region is that
emerging in the forward direction from this same slab. Clearly this 
approximation ignores a number of (unknown) geometrical factors relating to 
the spatial distribution of the scattering medium and our viewing 
orientation, but is adequate for the present purpose. Thus one set of spectral 
parameters apply to both the mtable and atable components. The three free 
parameters associated with the photoionized
gas are, log $\xi$, $N_{H_{scat}}$ and  $A_{em}$, the latter representing
the normalisation 
of the emission spectrum\footnote{See the {\sc xstar\/} manual for the
definition of this normalisation parameter (Kallman \& Krolik 1997)}.
In our modelling we have chosen not to
directly couple the scattered fraction of continuum flux 
(i.e. $A_{soft}$ divided by either $A_{ASCA}$ or $A_{Ginga}$) with  
$A_{em}$ (see section 4.2).

The results of fitting this photoionization model are given as Model 5 in 
Table 3 (and also in Table 2). This model gives a slightly 
better fit to the Mrk 3 spectrum than is the case for pure electron 
scattering plus a Raymond-Smith component (Model 4, Table 3). The form of the
spectrum in 0.3--3 keV band is shown in Figure 3(b).
The best-fitting model has log $\xi$ $ \approx 2.8$ and reproduces
the observational feature near 0.9 keV as recombination directly  
into the K-shell of a completely stripped oxygen atom, a process which has 
a threshold energy of 0.87 keV. In Fig. 3(b) this O VIII recombination
continuum feature is significantly broadened by the thermal energies of the 
free electrons in the $10^{6}$ K photoionized medium.

The derived slope of the intrinsic power-law continuum in Models 4 and 5  
remains relatively flat ($\alpha \approx 0.45$ and $ \approx 0.66$
respectively).
One possible explanation of the apparent hardness of the underlying 
continuum in Mrk 3 is that Compton reflection  makes a significant 
contribution to the observed flux above $\sim 6$ keV as appears
often to be the case for Seyfert 1 galaxies (Pounds et al. 1990).
We have investigated this possibility by including a Compton reflection 
component in the our spectral analysis; specifically we employ the XSPEC 
``pexrav'' model. We assume that the reflection component is subject 
to absorption in the cold gas column density represented by  
$N_H$. In pexrav the parameter $R$ represents the relative strength of the 
reflected signal, with a value of unity corresponding to reflection from a 
semi-infinite flat disk of cold material which is irradiated by an isotropic 
power-law continuum source. Here, for simplicity, we fix the inclination 
angle of the disk at i = 60$^{\circ}$ and assume the same
value of $R$ applies to the {\sl Ginga} and {\sl ASCA} observations.
The results of including reflection in
the photoionized scattering model are listed in Table 3 as 
Model 6. The best-fitting version of Model 6 requires $ R \sim$ 0.8,
(in the F-test the reduction in $\chi^{2}$ of $\sim$ 9 is significant at 
the $>$ 99 $\%$ level for one additional parameter).  Alternatively, when the
relative strength of the reflected signal is frozen at R = 1 and the
inclination angle of the disk allowed to vary, we obtain a best-fit value of i
= $71^{+13}_{-16}$ deg.  The inclusion of reflection in 
the spectral
model leads to a value for the spectral index of the hard continuum in Mrk 3
of $\alpha \approx 0.8$, close to the canonical value of $0.9$ measured
similarly in Seyfert 1 galaxies (Nandra \& Pounds 1994; Reynolds 1997; George
et al. 1998).

In our earlier discussion of the {\sl Ginga} spectra of Mrk 3 we noted
the evidence for additional line-of-sight absorption (over and above that 
represented by $N_H$) in the form of 
an edge feature near $\sim 8$ keV.  If attributable to iron-K edge
absorption, the implied ionization state of the absorbing gas is 
FeXVIII -- FeXXIII. This is rather close to the
state of the iron in the photoionized scattering region and suggests
a possible model in which the scattering zone extends not only directly
above the nuclear source (i.e. along the axial direction of the nuclear 
region) but also sufficiently close to the plane of the system to intercept 
our line of sight (the presumption being that we view the nucleus at fairly 
high inclination in the case of this ``classical'' Seyfert 2 galaxy).
This ``warm absorbing'' medium would presumably lie inside the colder medium
responsible for the bulk of the soft X-ray cut-off, a situation which is quite
plausible if the cold gas distribution is associated with a molecular torus 
in Mrk 3. We have investigated this twin absorbing-zone model by including 
a warm column density $N_{H_{warm}}$ in addition to the cold column $N_H$ in 
a revised version of Model 5. Again for simplicity we use the same mtable as 
described earlier and tie the ionization parameter of the warm absorber to 
that of the scattering medium. The results of this analysis are given in 
Table 3 as Model 7. We note that the fit is an improvement on both Models 5 
and 6. The inferred spectral index of the hard continuum in this 
``no-reflection'' scenario is $\alpha \approx 0.7 $;
a value which was considered typical of the underlying hard continuum in
Seyfert 1 galaxies prior to the realisation of the importance of the
Compton reflection process (e.g. Turner \& Pounds 1989). Finally we
note that the requirement for a warm-absorber is not diminished if we
add a reflection component to Model 7; however, we have not pursued this
model further given its complexity in relation to the quality 
of the available data.

\section{Discussion}

\subsection{Characteristics of the hard X-ray spectrum of Mrk 3}

Gross spectral variability in Mrk 3 might, in principle, be revealed 
as a mismatch between the {\sl Ginga}, {\sl ASCA} and {\sl ROSAT} spectral 
datasets. In fact we obtain a rather good fit to the composite spectrum 
without the need for any cross-instrument scaling factors (this also 
demonstrates that the instrument calibrations are not too far out of line). 
We are thus able to verify our initial assumption that the spectral 
form of Mrk 3, within the limitations of the available observations, is 
largely invariant. Specifically the soft X-ray spectrum of Mrk 3
(below $\sim 3$ keV) shows no significant change over the six week 
period separating the {\sl ROSAT} and {\sl ASCA} observations
and similarly the form of the spectral cut-off due to photoelectric 
absorption of the hard continuum flux remained 
relatively constant over the 3.5 year period between the {\sl Ginga} and 
{\sl ASCA} observations. In combination the {\sl Ginga} and {\sl ASCA} 
data give reasonably tight constraints on  the spectral slope of the hard 
continuum, $\alpha$, albeit in a model dependent way (as illustrated by 
the spread in values in Table 3). In practice we cannot exclude modest 
temporal variations in  $\alpha$ on the basis of the available
datasets. 

The two ``untied'' parameters in the spectral
fitting were the normalisation of the intrinsic power-law continuum 
and the flux of the iron K$_{\alpha}$ line.
The inferred absorption corrected flux of the hard X-ray 
source in the 2--10 keV band at the time of the {\sl Ginga} observations 
is 6.1 $\times 10^{-11} \rm~erg~cm^{-2}~s^{-1}$, whereas in the {\sl ASCA}
observations the value is 1.3$ \times 10^{-11} \rm~erg~cm^{-2}~s^{-1}$
(these and subsequent values quoted in this section
are based on Model 5). Thus variability in the 
continuum level by a factor $\sim 4.7$ is implied. As previously noted by 
Iwasawa et al. (1994), an observation by BBXRT (Marshall et al. 1991)
made roughly 14 months after the {\sl Ginga} observation gave an X-ray flux 
a factor 2 below the {\sl Ginga} level. 

The observed iron-line flux decreased by a factor of 1.8 between
the {\sl Ginga} and {\sl ASCA} observations, substantially less
than the corresponding continuum variation. Consequently the observed
equivalent width of the iron K$_{\alpha}$ line (measured relative 
to the {\it absorbed} continuum) shows an increase from 425 to 1110 eV 
over the same interval. 
The measured iron-line energy, corrected for a redshift 
$ z = 0.0137$, is $E_{K\alpha} = 6.38 \pm 0.03$ keV consistent with 
neutral iron.  
A possible origin for the iron K$_{\alpha}$ emission is in the fluorescence
of an extensive distribution of cold gas surrounding the nucleus of Mrk3, 
possible in the form of a torus. Our specific line of sight to the
the source of the hard X-ray continuum gives a column density 
$N_{H} = 7.7 \times 10^{23} \rm~cm^{-2}$, which if applicable over
$4\pi$ steradians implies  an iron-line equivalent width of $\sim 500$ eV
(Leahy \& Creighton 1993). Given the uncertainty as to whether the 
{\sl Ginga} or {\sl ASCA} continuum level is more representative of the 
average long-term luminosity of Mrk 3 and our lack of detailed
knowledge concerning both the coverage fraction and iron abundance of the 
fluorescing gas, it is plausible that the
bulk of the observed iron line emission originates in this fashion.  
However, Compton reflection from the surfaces of optically thick media, such 
as a putative accretion disk or the inner walls of sections of the torus, 
may also contribute to the 6.4 keV iron line. Our comparison of the 
{\sl Ginga} and {\sl ASCA} data demonstrates that we see hard X-ray 
variability down to $\sim 4.5$ keV (see Fig. 2) in Mrk 3 and suggests that 
we have a direct view of the power-law continuum (albeit with some losses due 
to the line-of-sight absorption) down to this energy.
In this case the predicted contribution to the equivalent width of the 
iron line measured with respect to the observed continuum is
$ ^{<}_{\sim} 300$ eV (e.g. George \& Fabian 1991; Krolik \& $\dot{\rm Z}$ycki
1994; Ghisellini, Haardt \& Matt 1994). The observed decline of the iron line 
flux between the {\sl Ginga} and {\sl ASCA} observations demonstrates that
this line does show some response on timescales of a few years, which
in turn requires at least one of the line emitting regions to have
a dimension not much more than about a light year. It would seem more likely 
that this scale size corresponds to an accretion disk or inner (optically 
thick) cloud structure rather than the inner extent of the putative 
molecular torus. In any event, milliarcsec resolution will be required
to image this spatial structure.

The profile of the iron K${_\alpha}$ line in Mrk 3 is probably complex. For 
example, in the archetypal Seyfert 2 galaxy, NGC 1068, up to four line 
components have been spectrally resolved in the
6--7 keV energy band in {\sl ASCA} observations (with a summed line
photon flux which is $\sim 3$ times higher than is the case for Mrk 3)
(Ueno et al. 1994; Iwasawa, Fabian \& Matt 1997).
The identified line features include a 6.4 keV line, probable
He-like and H-like iron K$_{\alpha}$ components and also the hint
of a ``Compton shoulder'' feature on the 6.4 keV line
(Iwasawa, Fabian \& Matt 1997). The He-like and H-like lines
have been attributed to a warm scattering region 
(Marshall et al. 1993; Matt, Brandt \& Fabian 1996)
which is probably smaller in extent than the optical/UV mirror
in this source (Iwasawa, Fabian \& Matt 1997).
Some evidence for underlying complexity in the iron K${_\alpha}$ 
line is provided, in the case of Mrk 3, by the fact that in the spectral 
fitting we minimize $\chi^{2}$ when we fix the width of the 6.4 keV
at a measureable value, namely $\sigma_{K\alpha}= 0.1$ keV.
(In Model 5, Table 3 the $\chi^{2}$ increased by $\sim 6$ when we
reduced the line-width by a factor of 2). The origin of such
broadening could be due to relativistic effects in an accretion 
disk (e.g. Tanaka et al. 1995; Yaqoob et al. 1995; Nandra et al. 1997)
or the superposition of different line components. For example,
our photoionized scatterer model predicts up to 20$\%$ of the iron 
K$_{\alpha}$ flux observed by {\sl ASCA} will be in the form of 
Fe XVIII - XXV, although this does not account for all of the apparent line
broadening in Mrk 3.

Although previous studies have suggested that the power-law continuum in 
Mrk 3 is unusually hard (e.g. Awaki et al. 1990; Awaki et al. 1991;
Smith \& Done 1996), as discussed in section 3.3, we obtain a value
for the energy spectral index, $\alpha \sim 0.8$, which is well within
the range of values seen in Seyfert 1 galaxies
(Nandra \& Pounds 1994; Reynolds 1997; George et al. 1998).
Even without reflection we require $\alpha \sim 0.7$ if we include
an inner warm-absorber so as to reproduce the $\sim 8$ keV absorption
edge apparent in the {\sl Ginga} spectra. Our best fitting reflection
model (Model 6, Table 3) requires a reflection component of
only modest strength (i.e. $R \sim 0.8$), which contributes significantly
to the observed spectrum only above 10 keV. In a recent study
Turner et al. (1997b) suggest an alternative description of the 
hard X-ray spectrum of Mrk 3 in which the reflection component is 
effectively free of absorption and contributes the bulk of the observed
flux between 3--10 keV. In this model the intrinsic power-law 
continuum suffers line-of-sight absorption by a cold gas 
column density $N_H \approx 1.3 \times 10^{24} \rm~cm^{-2}$ (roughly 
a factor of 2 higher than reported here) and is thus strongly suppressed
below $\sim 6 $ keV. The current analysis demonstrates that continuum 
variability is observed in Mrk 3 down to $\sim 4.5$ keV with little
evidence for a changing spectral form.  If the continuum above 4.5 keV 
represents the combination of reflection and a very heavily obscured 
power-law component, as in the Turner et al model, this would require the 
reflection component to track the power-law at a rate at least comparable 
to the intrinsic variability timescale (i.e. a timescale shorter than
$\sim$ a year). One advantage of the Turner et al. (1997a) model is that
the large iron-line equivalent width measured by {\sl ASCA} can be readily
explained; however, as discussed earlier, when the continuum variability is 
taken into account, the {\sl ASCA} measurement can also be reconciled with
the current model in which the power-law component is observed directly
down to $\sim 4.5$ keV.

\subsection{The nature of the soft X-ray emission in Mrk~3}

The soft X-ray excess in the spectrum of Mrk 3 (which dominates the spectrum
below $\sim 3$ keV) has been modelled either as a blend of an electron
scattered power-law continuum with a soft thermal component  (Model 4) or an 
absorbed continuum scattered from a hot photoionized medium, with line 
emission from the same region superimposed (Model 5). We now 
consider each of these scenarios in turn.

In Model 4, a substantial fraction of the soft X-ray luminosity
is explained as emission from a Raymond-Smith plasma at a 
temperature of $\sim 0.7$ keV. A possible origin of this emission is in 
shock heated gas associated with a nuclear starburst in Mrk 3.  The ROSAT HRI
measurements of Morse et al. (1995) constrain the X-ray emitting region in
Mrk 3 to be $< 4$ kpc, and thus do not exclude the presence of a nuclear 
starburst. In a recent study of the {\sl ASCA} spectra of a large sample of
Seyfert 2/Narrow Emission Line Galaxies, Turner at al. (1997a) used
far infra-red luminosities to estimate an upper limit to the fraction of 
the soft X-ray flux that could arise from starburst activity
(following the method first discussed by David, Jones \& Forman 1992).
On this basis Turner et al. (1997a) concluded that in Mrk 3 no more than 
6 $\times 10^{40} \rm~erg~s^{-1}$ in the 0.5--4.5 keV band could be thermal
emission powered by starbursts, which represents only 5$\%$ of the 
total 0.5--4.5 keV luminosity of the source. The thermal component in 
Model 4 has a 0.5--4.5 keV luminosity
of $2.1 \times 10^{41} \rm~erg~s^{-1}$ (corrected for Galactic absorption),
a factor 3.5 higher than the prediction. However the scatter 
in the correlation between the soft X-ray and far-infrared luminosity 
in the sample of normal and starburst galaxies studied
by David et al. (1992) is of the same order; hence we conclude that 
a starburst origin for the thermal emission is not excluded.  
An alternative possibility is that the thermal emission arises in gas which 
is energised by the radio jet in Mrk 3 and possibly associated with the
small-scale bar ($0.35'' \times 1''$) of continuum emission seen in the 
recent {\sl HST} observation of Mrk 3 (Capetti et al. 1995).

The Raymond-Smith component contributes 23\% of the observed 0.3--3 keV 
emission (see Fig. 3(a)) with the electron scattered component
accounting for the remainder. The scattered fraction, $f_{scat}$, 
of the hard continuum flux is  $A_{soft}$/$A_{Ginga} \approx 2.7\%$  
(or as high as $\approx 12\%$ if the level of the hard continuum
in the {\sl ASCA} observation is more representative of the time averaged
behaviour of the source). For comparison Schmidt and Miller (1985) estimate 
that the scattered fraction is $<10\%$ based on the degree of polarization 
in the optical spectrum of Mrk 3. 
The column density of the scattering medium can be estimated
from the fraction $f_{scat}$, since $f_{scat} =$ $\Omega\over{4\pi}$
$\tau_{s}$  where ${\Omega}$ is the solid angle subtended
by the scatterer (as viewed from the continuum source) and $\tau_{s}$
is the electron scattering optical depth given by $\tau_{s} = 1.2
\sigma_{T}N_{Hscat}$. If we use the lower of the two estimates for
the scattered fraction and take  $\Omega\over{4\pi}$ $= 0.25$ then we
obtain $N_{Hscat} \sim 1.3 \times 10^{23} \rm~cm^{-2}$.  

We now turn to our alternative description of the soft X-ray spectrum of 
Mrk 3. In Model 5 the scattering region is not completely
photoionized and thus emprints both absorption and emission features
on the emergent spectrum. The value for the 
column density,  $N_{H_{scat}} \sim 1.7 \times 10^{23}\rm~cm^{-2}$,
obtained from the spectral fitting is consistent with the previous
estimate of a scattered fraction of $\sim 3\%$. 
The ionization parameter of log $\xi$ $\sim$ 2.8 corresponds to iron 
ionization states in the range Fe XVIII - XXV and oxygen either in a 
hydrogenic or fully ionized form (O VIII - IX) (Kallman \& McCray 1982).
Although this is on the high side of the range, it is not totally
atypical of the degree of ionization inferred for the warm absorbers 
seen in Seyfert 1 galaxies (Reynolds 1997; George et al. 1998).
We have assumed that the plasma temperature is $10^{6}$ K, which is
hotter than the warm scattering which acts as an optical/UV
mirror in NGC 1068 (Miller, Goodrich \& Mathews 1991). This is also
above the limit of thermally stability in the photoionization models
considered by Netzer (1993, 1996). However, the plasma state is similar to 
the case considered by Krolik \& Kriss (1995) (their model a), where the 
scattering region is identified as an X-ray heated wind which is in 
ionization equilibrium but not necessarily radiative equilibrium. 
Empirically we find that reducing the plasma temperature to either 
$10^{5.5}$ or $10^{5}$ K produces a set of photoionization models which 
provide a somewhat poorer best fit to the observational data, particularly 
near the 0.9 keV feature (see below), although many of the details of 
the spectral model are similar for plasma temperatures in the 
range $10^{5-6}$ K. Tran (1995) notes that the observed FWHM of the 
polarized broad H$\beta$ in Mrk 3 is $6000 \rm~km~s^{-1}$, implying
that the temperature of the electrons which scatter the optical
light is less than $4 \times 10^{5}$ K. The confirmation of
a $10^{6}$ K component through X-ray spectroscopy might thus require
thermal stratification of the scattering medium.

The observed feature in the {\sl ASCA} spectrum near $\sim$ 0.9 keV 
is explained as recombination directly into the K-shell of fully
ionized oxygen atoms; the resulting O VIII recombination continuum
is broadened by the thermal energies of the electrons in the 
$10^{6}$ K plasma.  Fig. 3(b) (and also Fig 3(a)) also shows 
a prominent line feature at 0.65 keV due to the O VIII Ly${_\alpha}$ line,
with an equivalent width of $\sim 30$ eV. The observational evidence
for this feature is marginal given that when we remove the line 
from the spectral model the result is a change in the $\chi^{2}$ of
less than 1. Similarly
the model also predicts weak spectral features in the energy band
from 1.8--3 keV due to silicon and sulphur which are currently at the 
limit of detectability with {\sl ASCA}.  

In a recent paper, Comastri et al. (1997) considered the {\sl ASCA} spectrum
of the Seyfert 2 galaxy NGC 4507 in which a line-like feature is again 
observed near 0.9 keV. These authors attribute this feature in NGC 4507 to 
a helium-like Neon line at a rest energy of 0.92 keV. In this case OVIII 
recombination at 0.87 keV is formally inconsistent (at the $3\sigma$ level) 
with the observed line energy. However, Comastri et al. (1997) note that in 
a $10^{6}$ K plasma, the broadening of the OVIII feature would have the 
effect of shifting the centroid energy to higher energy and therefore could 
be made consistent with the NGC 4507 observations. In fact we find, from a
preliminary analysis of the relevant datasets, that the 
composite {\sl Ginga}, {\sl ASCA} and {\sl ROSAT} spectrum of NGC 4507 can be 
well described by Model 5, with parameter values quite similar to those 
pertaining in Mrk 3.

The scale of the X-ray scattering region in Mrk 3 can be estimated
from the derived value of the ionization parameter and the model 
assumptions. We obtain a radius $r = 5 \times 10^{17}n_{6}^{-0.5}\rm~cm$
(where $n_{6}$ is the density in units of $n = 10^{6} \rm~hydrogen~nuclei~
cm^{-3}$). The measured column density through the scattering region 
leads to an estimate of the thickness of the scatterer of 
$l = 1.7 \times 10^{17} n_{6}^{-1.0}\rm~cm$. Then if we require 
$ l~^{<}_{\sim}~r $, it follows that $n_{6}$ $_{\sim}^{>}~0.1$. 
The implied scale of the scattering  region ($\sim$ a light year) is 
comparable to the inner extent of the cold fluorescing gas as inferred 
from the variability in the iron-line flux.

Finally we note the complication that the predicted normalisation
of the emission spectrum is a factor $\sim 2$ times higher than 
the best fitting value. This apparent preference for scattered continuum 
as opposed to intrinsic emission from the photoionized region
may reflect the fact that our representation of the spectral form of the 
incident continuum in the soft X-ray band, as a straight downward 
extrapolation of the hard power-law, may not be adequate. For example, the 
(hidden) intrinsic nuclear spectrum in Mrk 3 may show an upturn somewhere 
in the 0.3--3 keV band similar to that seen in some unobscured Seyfert 1 
nuclei (e.g. George et al. 1998).

\section{Conclusions}
The main results of our analysis of the composite {\sl Ginga},
{\sl ASCA} and {\sl ROSAT} spectrum of Mrk 3 are as follows: \\
(i) The broad-band 0.1--30 keV X-ray spectrum is rather stable with
only the level of the hard X-ray continuum and the iron K$_{\alpha}$ 
line-flux showing clear evidence of variability on a timescale of years. \\
(ii) The iron K$_{\alpha}$ line energy is consistent with neutral iron.
A component of this line emission may originate in the fluorescence of an
extensive cold gas distribution, which has a column density  $N_{H} \sim 7 
\times 10^{23} \rm~cm^{-2}$ along our line of sight to the active
nucleus; it is plausible that this gas is  associated with a 
molecular torus in Mrk 3. Photoionization of the inner reaches of this
medium may give rise to the ``warm'' absorption edge observed near 8 keV.
The observed changes in the line equivalent width demonstrate that the iron 
line has a delayed response with respect to the continuum variations but, 
nevertheless, that iron-line reprocessing, possibly in an accretion disk, 
occurs within $\sim$ a  light-year of the active nucleus in Mrk 3. \\
(iii) The measured iron line width of $\sigma_{K\alpha} \approx 0.1$ keV 
provides a hint of underlying complexity in composition of the iron 
K$_{\alpha}$ line, possibly similar to the situation in NGC 1068. \\
(iv) The inferred slope of the intrinsic hard X-ray continuum is model 
dependent with either a modest reflection component or the inclusion of an 
inner warm-absorber component both serving to steepen the derived
spectral index. The continuum  appears {\it not} to be anomalously 
flat compared to the norm for Seyfert 1 galaxies. \\
(v) The soft X-ray spectrum below $\sim 3$ keV is quite complex and may 
represent a blend of two or more separate components. A possible 
scenario is that the emission arises from the scattering of the
nuclear continuum by a hot photoionized plasma which emprints both 
absorption and emission features on the emergent spectrum. In this setting
the spectral feature present in the {\sl ASCA} spectrum near 0.9 keV 
is identified with OVIII recombination.

Clearly there is much to be learnt about Mrk 3 and other Seyfert 2 galaxies
from X-ray spectral observations affording an appropriate combination of high 
spectral resolution, high sensitivity and broad-bandwidth coverage.
Fortunately future missions such as {\sl AXAF}, {\sl XMM} and {\sl ASTRO-E} 
will provide the opportunity to fully exploit this window on the 
Seyfert phenomenon.

\section{ACKNOWLEDGEMENTS}

The authors would like to thank Tim Kallman for his advice on using
\xstar for this analysis. RGG acknowledges support from PPARC in the form
of a research studentship. CD acknowledges support from a PPARC 
advanced fellowship.  The X-ray spectral data used in this work were obtained
from the Leicester Data Archive Centre (LEDAS) and the HEASARC facility at
Goddard Space Flight Centre, USA.

\pagebreak


\newpage

\setlength{\parindent}{0pt}

{\bf Figure Captions} \\

{\bf Figure 1.} Top panel: The {\sl Ginga} LAC top-layer (triangles) and 
mid-layer (circles) count rate spectra for Mrk 3. The solid line corresponds 
to the best fitting version of Model 1. Bottom panel: The spectral fitting 
residuals to the {\sl Ginga} data for Mrk 3 (scaled in terms of the 
measurement errors).\\

{\bf Figure 2.} Top panel: The unfolded composite spectrum of Mrk 3 based on 
the best-fitting version of Model 3. Bottom panel:
The spectral fitting residuals to the composite spectrum for
Mrk 3. The data points correspond to {\sl Ginga} top-layer (circles),
{\sl Ginga} mid-layer (squares), {\sl ASCA} SIS-0 (crosses) and {\sl ROSAT}
PSPC (triangles). For clarity only the {\sl ASCA} SIS-0 points are shown.\\

{\bf Figure 3.} Modelling of the soft X-ray (0.3--3 keV) spectrum of Mrk 3.
(a) As a blend of a soft (scattered) power-law continuum with a 
$\sim 0.7$ keV Raymond-Smith component. The spectrum from 
0.7--0.9 keV is dominated by O, Fe (L shell)
and Ne lines. (b) As an absorbed continuum
scattered from a hot photoionized medium plus line emission from 
the same region. The feature near 0.9 keV is due to O VIII recombination 
and that at 0.65 keV corresponds to the O VIII L$_{\alpha}$ line. \\

\newpage

\begin{table}
 
\caption{\large Spectral fitting of the {\sl Ginga} data alone}

\large

\begin{tabular}{lcc}
 
& & \\
Parameter & Model 1 & Model 2 \\

& & \\
$\alpha$ & 0.37$^{+0.15}_{-0.14}$& 0.65$^{+0.20}_{-0.19}$ \\ 
$A_{Ginga}$~$^{a}$& 0.6$^{+0.3}_{-0.2}$ & 1.5$^{+1.1}_{-0.6}$ \\ 
$N_H$~$^{b}$ & 58$\pm 7$ & 69$\pm 9$ \\ 
& & \\
$E_{\rm K_{\alpha}}$~$^{c}$& 6.39$\pm 0.10$ & 6.33 $^{+0.19}_{-0.21}$ \\ 
$A_{\rm K_{\alpha}}$~$^{d}$& 7.9$\pm 1.4$ & $4.0^{+2.2}_{-2.4}$ \\ 
& & \\
$E_{edge}$~$^{c}$ & - & 8.1$^{+0.4}_{-0.5}$ \\
$\tau_{edge}$          & - & 0.24$\pm 0.10$ \\
& & \\
$\chi^{2}$& 56.0  & 42.1 \\
d.o.f.    & 56	   & 54  \\ 
& & \\
\end{tabular}

\raggedright

$^{a}$ $10^{-2}$ photon cm$^{-2}$ s$^{-1}$ keV$^{-1}$. \\
$^{b}$ $10^{22}$ cm$^{-2}$. \\
$^{c}$ keV. \\
$^{d}$ $10^{-5}$ photon cm$^{-2}$ s$^{-1}$.\\

\end{table}

\begin{table}

\caption{\large Spectral fitting of the {\sl ASCA} SIS/GIS data alone}

\large

\begin{tabular}{lccc}

& & & \\
Parameter	 & Model 3  & Model 4  & Model 5 \\

& & & \\
$\alpha$ & 1.12$\pm 0.10$ & 0.56$\pm 0.16$ & 0.91$^{+0.11}_{-0.12}$ \\ 
$A_{ASCA}$~$^{a}$& 0.45$^{+0.12}_{-0.09}$ & 0.14$^{+0.06}_{-0.05}$ & 0.32$^{+0.11}_{-0.08}$ \\
$N_{H_{1}}$~$^{b}$ & 41.4$^{+6.0}_{-5.3}$ & 44.0$^{+8.2}_{-6.8}$ & 48.0$^{+8.7}_{-7.1}$ \\
& & & \\
$E_{K_{\alpha}}$~$^{c}$& 6.31$\pm 0.03$ & 6.30$\pm 0.03$ & 6.30$\pm 0.03$ \\
$A_{K_{\alpha}}$~$^{d}$& 4.2$\pm 0.6$ & 4.2$\pm 0.6$ & 4.1$\pm 0.6$ \\
& & & \\
$A_{soft}$~$^{a}$ & 0.035$\pm 0.002$ & 0.022$\pm 0.003$ & 0.037$^{+0.006}_{-0.005}$ \\
& & & \\
$kT_{Ray}$~$^{c}$ & -- & 0.74$ \pm 0.06$ & -- \\
$A_{Ray}$~$^{a}$ & -- & 0.008$^{+0.002}_{-0.002}$ & -- \\
& & & \\
log $\xi$ & -- & -- & 2.42$^{+0.27}_{-0.24}$ \\
$N_{H_{scat}}$~$^{b}$ & -- & -- & 6$^{+5}_{-3}$ \\
$A_{em}$~$^{e}$ & -- & -- & 1.1$\pm 0.4$ \\
& & & \\
$\chi^{2}$& 390.0  & 321.1   & 318.8 \\
d.o.f.    & 300	   & 298     & 297   \\ 
& & & \\ 
\end{tabular}

\raggedright

$^{a}$ $10^{-2}$ photon cm$^{-2}$ s$^{-1}$ keV$^{-1}$. \\
$^{b}$ $10^{22}$ cm$^{-2}$. \\
$^{c}$ keV. \\
$^{d}$ $10^{-5}$ photon cm$^{-2}$ s$^{-1}$.\\
$^{e}$ $10^{-12}$ photon cm$^{-2}$ s$^{-1}$.\\

\end{table}

\pagebreak

\begin{table*}

\caption{\large Spectral fitting of the composite {\sl Ginga}, {\sl ASCA} and
{\sl ROSAT} spectra for Mrk 3}

\large

\begin{tabular}{lccccc}

& & & & & \\
Parameter	& Model 3  & Model 4 & Model 5  & Model 6 & Model 7\\
& & & & & \\
$\alpha$ & 0.88$\pm 0.06$ & 0.45$\pm 0.08$ & 0.66$\pm 0.07$ &
0.80$^{+0.07}_{-0.11}$ & 0.69$\pm 0.07$ \\ 
$A_{Ginga}$~$^{a}$ & 2.58$^{+0.47}_{-0.40}$& 0.77$^{+0.20}_{-0.16}$	&
1.39$^{+0.29}_{-0.25}$	& 1.62$^{+0.29}_{-0.33}$ & 1.63$^{+0.26}_{-0.28}$ \\
$A_{ASCA}$~$^{a}$ & 0.64$^{+0.14}_{-0.12}$& 0.17$^{+0.06}_{-0.04}$	&
0.29$^{+0.08}_{-0.07}$	& 0.36$^{+0.08}_{-0.09}$ & 0.34$^{+0.08}_{-0.07}$ \\
$N_{H}$~$^{b}$ 	& 83.4$^{+3.6}_{-3.5}$ 	& 66.4$\pm 4.1$		& 77.0$\pm
4.0$ & 71.4$^{+4.6}_{-4.8}$ & 66.8$^{+6.6}_{-3.7}$ \\
& & & & &\\
$E_{K_{\alpha}}$~$^{c}$	& 6.29$\pm 0.03$ & 6.30$\pm 0.03$ &6.29$\pm 0.03$
& 6.29$\pm 0.03$ & 6.28$\pm 0.03$ \\
$A_{K\alpha}$$(Ginga)$~$^{d}$& 6.0$^{+1.4}_{-1.5}$ & 7.6$\pm 1.4$ &7.0$\pm
1.4$  & 5.7$\pm 1.5$ & 5.2$\pm 1.5$  \\
$A_{K\alpha}$ $(ASCA)$~$^{d}$	& 3.8$\pm 0.6$ 	& 4.1$\pm 0.6$ 	&
3.9$\pm 0.6$ & 3.8$^{+0.6}_{-0.5}$ & 3.8$\pm 0.6$  \\
& & & & & \\
$A_{soft}$~$^{a}$ & 0.033$\pm 0.001$ 	& 0.021$\pm 0.002$ 		&
0.038$\pm 0.004$ 	& 0.039$\pm 0.004$ & 0.039$ ^{+0.005}_{-0.004}$ \\
& & & & & \\
$kT_{Ray}$~$^{c}$ & -- 	& 0.69$^{+0.07}_{-0.05}$	& -- 	& -- & -- \\
$A_{Ray}$~$^{a}$ & -- 	& 0.0085$^{+0.0018}_{-0.0011}$ & -- & -- & -- \\
& & & & & \\
log $\xi$ & -- & -- & 2.81$\pm 0.13$ & 2.61$^{+0.23}_{-0.08}$ & 2.78$^{+0.12}_{-0.19}$	 \\
$N_{H_{scat}}$~$^{b}$ 	& -- 	& -- & 17$^{+6}_{-5}$ 	& 10$^{+7}_{-3} $ &
16$^{+5}_{-6}$	\\
$A_{em}$~$^{e}$ & -- & -- & 0.9$\pm 0.2$ & 0.9$\pm 0.2$ & 0.9$\pm 0.2$  \\
& & & & & \\
$N_{H_{warm}}$~$^{b}$ & -- & -- & -- & -- & 39$^{+1^{*}}_{-18}$ \\
& & & & & \\
& & & & & \\
$R$	& --	&-- &--	& 0.8$\pm 0.5$ & -- \\
& & & & & \\
$\chi^{2}$	& 581.0  & 417.6 & 405.6 & 397.0 & 391.2 \\
d.o.f.    	& 378	 & 376 	 & 375   & 374   & 374   \\
& & & & & \\
\end{tabular}

\raggedright

$^{a}$ $10^{-2}$ photon cm$^{-2}$ s$^{-1}$ keV$^{-1}$. \\
$^{b}$ $10^{22}$ cm$^{-2}$. \\
$^{c}$ keV. \\
$^{d}$ $10^{-5}$ photon cm$^{-2}$ s$^{-1}$.\\
$^{e}$ $10^{-12}$ photon cm$^{-2}$ s$^{-1}$.\\
$^{*}$ reached upper limit of model. \\
\end{table*}

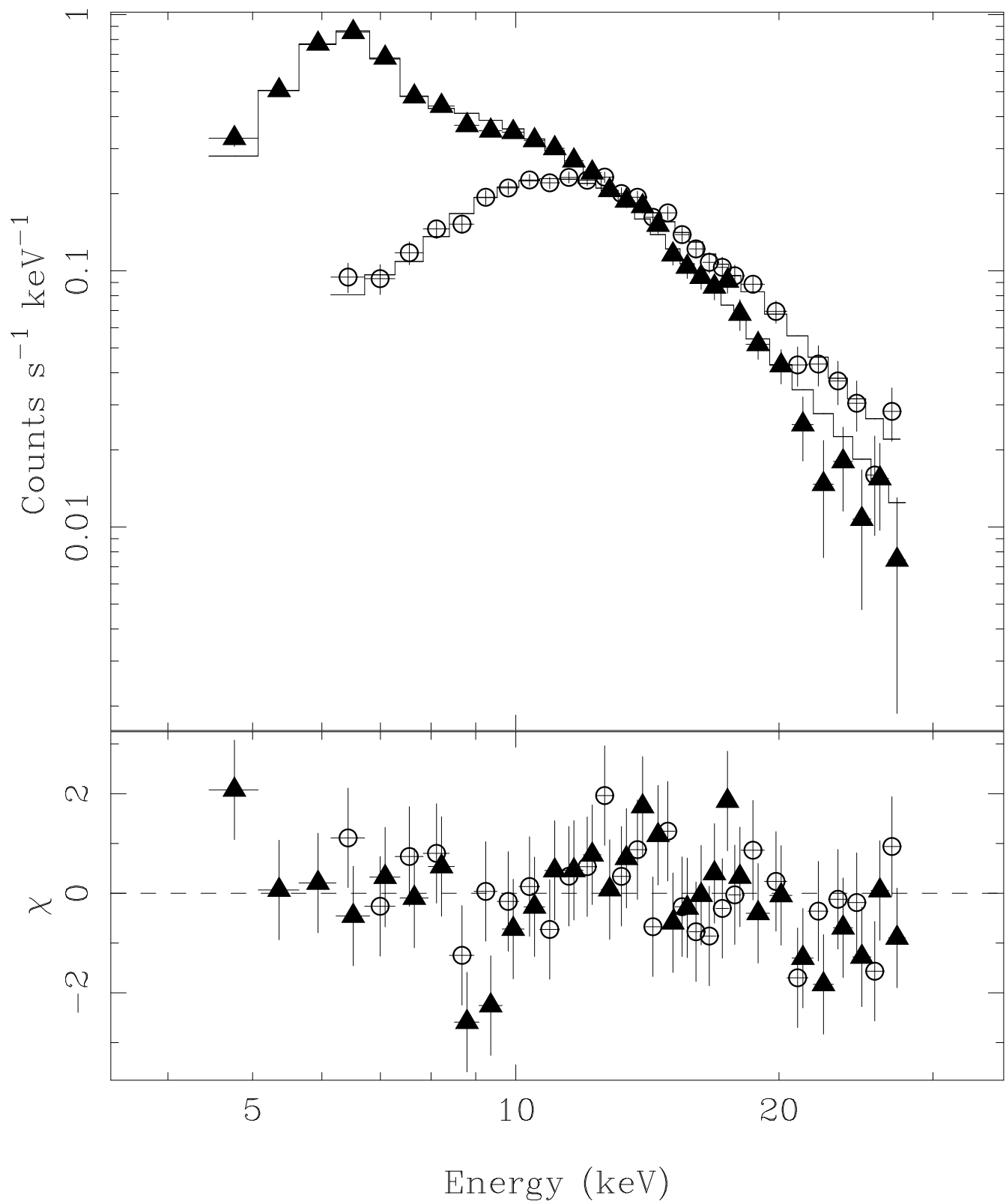
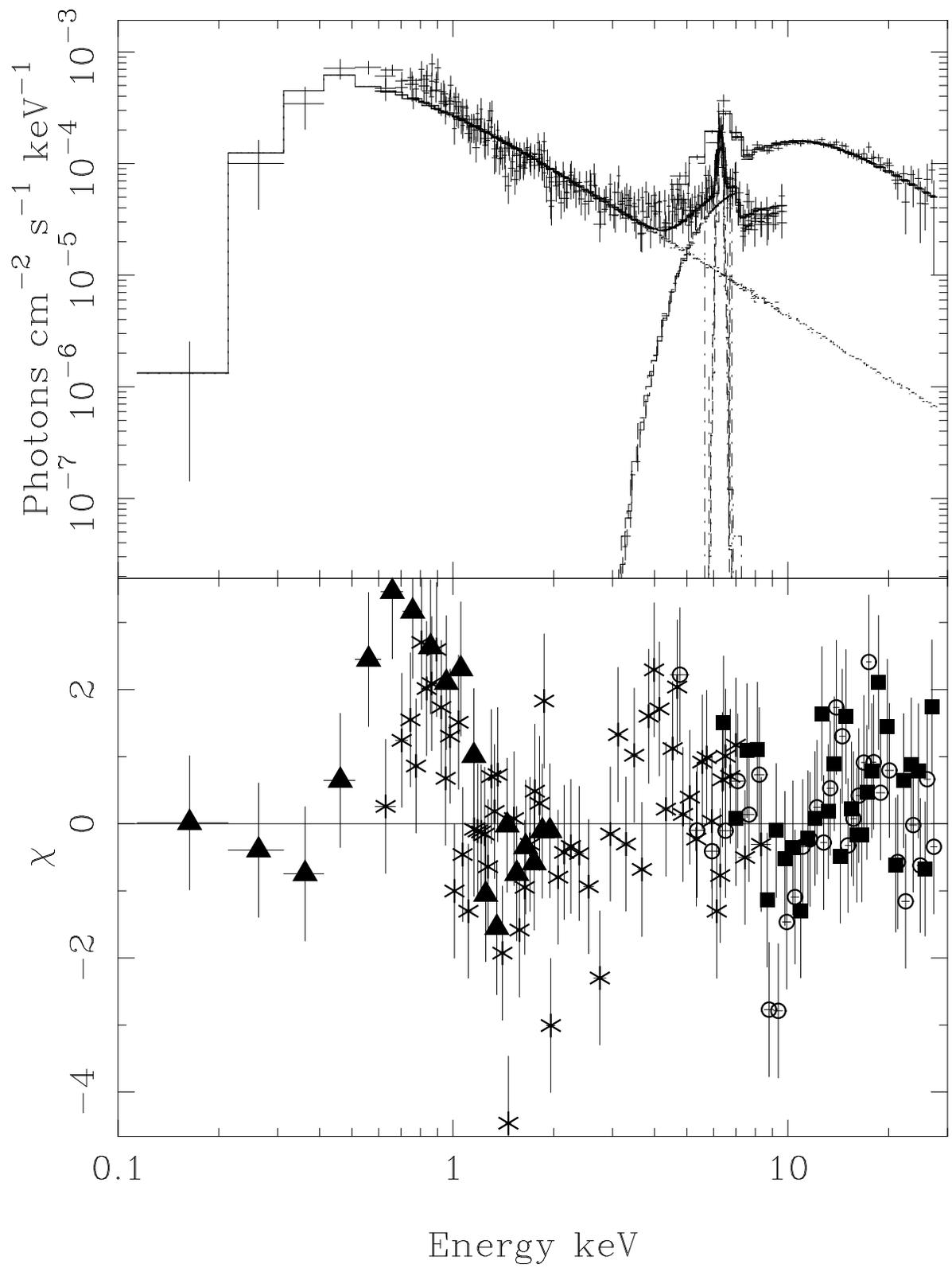
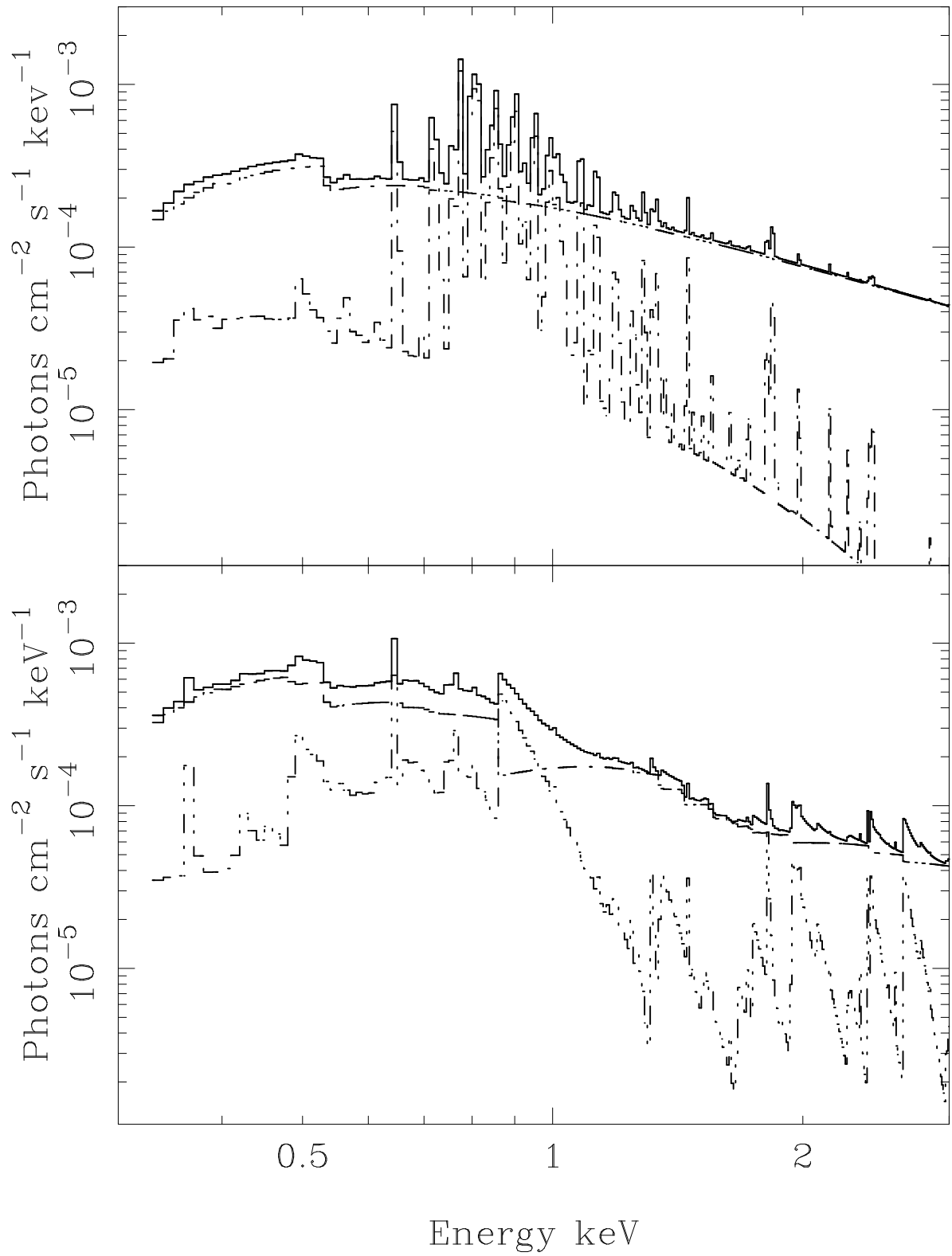
\end{document}